\documentclass[11pt]{article}
\usepackage{graphicx}
\usepackage{amsmath}
\usepackage{amssymb}
\usepackage{fancyvrb}

		\title{Optimising the reliability that can be claimed for a software-based system based on failure-free tests of its components}
				\author{Peter Bishop, Andrey Povyakalo\\City, University of London}
				\date{~}

\begin{document}
	
		
\maketitle

\begin{abstract}
	This short paper describes a numerical method for optimising the conservative confidence bound on the reliability of a system based on tests of its individual components. This is an alternative to the algorithmic approaches identified in \cite{bishop2020conservative}. For a given maximum number of component tests, the numerical method can derive an optimal test plan for any arbitrary system structure. 
	
	The optimisation method is based on linear programming which is more efficient that the alternative integer programming. In addition, the optimisation process need only be performed once for any given system structure as the solution can be re-used to compute an optimal integer test plan for a different maximum number of component tests.
	
	This approach might have broader application to other optimisation problems that are normally implemented using integer programming methods.
~

\end{abstract}



\section{Introduction}

Statistical testing \cite{ehrenberger1985statistical,parnas1991assessment,may1995} provides a direct estimate of the software probability of failure on demand ({\em pfd}) of a demand-based system to some confidence bound, and it is recommended in functional safety standards such as IEC 61508  \cite{iec61508:2010}. The standard approach to deriving a confidence bound on the {\em pfd} of a software-based system is to perform statistical testing on the whole system as a ``black-box''. In practice, performing tests on the entire system may be infeasible for logistical reasons, such as lack of availability of all component subsystems at the same time during implementation.

To address this issue, a general method was developed for deriving a conservative confidence bound  based on independent statistical tests applied to individual software-based components within the system \cite{bishop2020conservative}. The approach is completely general -- it can be used to derive a conservative {\em pfd} bound for any system architecture (represented by a structure function) for a given component test plan.

The choice of component test plan affects the {\em pfd} bound that can be achieved. The paper showed that for symmetrical architectures (like $n$ out of $m$ vote structures), an even split of tests between components always produces the optimal {\em pfd} bound (regardless of whether the software components are diverse or identical). 

Deriving an optimal test plan for arbitrary, asymmetric structures proved to be more of a challenge. Two sub-optimal test plan strategies were identified that are optimal for some asymmetric structures -- but not in general.

This paper presents an alternative to the test plan algorithms described in to \cite{bishop2020conservative} that derives an optimal test plan using linear programming. We first summarize the main elements of the theory presented in \cite{bishop2020conservative}, and then present our alternative method for generating an optimal test plan using numerical methods.

\section{Confidence Bounds from Component Tests}

Failure-free testing over $m$ individual components can be characterized by a test plan vector
\begin{equation}
\mathbf{n} =({n_1}, {n_2}, \dots, {n_m})'
\end{equation}
where $m$ is a number of components, $n_j$ is the number of (failure-free) tests for component $j$, and the total number of tests is
\begin{equation}
N = \sum_{j=1}^{m} n_j .
\end{equation}

To characterize the fault tolerance capability of a system architecture, we define $\mathbf{x} = ( x_1, x_2, \dots, x_m)'$ as a random binary vector of indicators of component failure. If component $j$ fails, $x_j=1$ and $x_j=0$ otherwise. 

The failure-proneness of the overall system is represented by a structure function $\varphi(\mathbf x)$, 
where $\varphi(\mathbf x)=1$ if the system fails for a given combination of component failures and successes $\mathbf x$. Such a system state is known as a {\em cutset}. 

Table \ref{tbl:2oo3_structure} shows the states for a 2 out of 3 (2oo3) vote structure where two or more component failures will result in system failure (i.e.~where $\varphi(\mathbf x)=1$), e.g. in state $\mathbf x_4$, failure of components $c_1$ and $c_2$ causes system failure.

\renewcommand{\arraystretch}{1.1}

\begin{table}[ht]
		\caption{Example 2oo3 vote structure function}
\label{tbl:2oo3_structure}	\begin{center}
		\begin{tabular}{|c||c|c|c||c|}
			\hline  Component  &   $c_1$  &  $c_2$& $c_3$ & 	\\
			\cline{1-4}   State $\mathbf x$ & $x_1 $  & $x_2$  & $x_3$  & $\varphi(\mathbf x)$\\ 
			\hline   $\mathbf x_0$ & 0 & 0 & 0 & 0 \\ 
			\hline   $\mathbf x_1$ & 1 & 0 & 0 & 0 \\
			\hline   $\mathbf x_2$ & 0 & 1 & 0 & 0 \\
			\hline   $\mathbf x_3$ & 0 & 0 & 1 & 0 \\
			\hline   $\mathbf x_4$ & 1 & 1 & 0 & 1  \\ 
			\hline   $\mathbf x_5$ & 1 & 0 & 1 & 1  \\
			\hline   $\mathbf x_6$ & 0 & 1 & 1 & 1 \\
			\hline   $\mathbf x_7$ & 1 & 1 & 1 & 1 \\			
\hline
		\end{tabular} 

	\end{center}
\end{table} 

A general proof given in \cite{bishop2020conservative} shows that, for any structure $\varphi(\mathbf \cdot)$, the upper confidence bound, $q_s$, for the system {\em pfd} can be conservatively approximated as
\begin{eqnarray}\label{eqn:high_approx}
&& q_s \le \min \left(\frac{\ln(1/\alpha)}{N_{min}},1\right)
\end{eqnarray}
where $N_{min}$ is the smallest total number of component tests in a cutset, i.e. 
\begin{eqnarray}\label{eqn:n_min}
&& N_{min} = \min_{\forall \mathbf x : \varphi(\mathbf x)=1}(\mathbf{n \cdot x})
\end{eqnarray}
where $\mathbf{n \cdot x}$ is the scalar product of the two vectors, i.e. $\sum_{j=1}^m n_j x_{j}$. 
For example, for the case where $\mathbf x = \mathbf x_4$ in Table \ref{tbl:2oo3_structure}, the scalar product will be \[1.n_1+1.n_2+0.n_3 = n_1 +n_2\]

For symmetrical structures, the optimal test plan is simple -- the $N$ tests are apportioned equally between the $m$ components, e.g., in the 2oo3 vote structure, each component is assigned $N/3$ tests so $N_{min}=2N/3$. 

It proved to be more difficult to identify the optimal test plan for arbitrary asymmetric structures. It was shown in \cite{bishop2020conservative}, that for any structure, the optimum test plan would always be able to achieve: 
\begin{eqnarray}\label{eqn:plan_lim}
&& N_{min} \ge \frac{N}{P}
\end{eqnarray}
where $P$ is the length (number of operational components) of the shortest success path. For example, in a 2oo3 vote structure, $P=2$ because we need at least two working components for correct system operation.

Two test plan strategies were identified in \cite{bishop2020conservative} that are optimal for some asymmetric structures -- but not in general. For example, one strategy assigned the $N$ tests equally to the $P$ components on a single shortest path. In the 2oo3 example, where $P=2$, this would mean assigning $N/2$ tests to, say, $c_2$ and $c_3$, and zero to $c_1$. This allocation results in $N_{min}=N/2$ which is clearly worse than the optimal value of $N_{min}=2N/3$.   

While further test plan allocation algorithms were examined, it was always possible to identify a counter-example structure where the allocation would be sub-optimal.

The alternative approach is to derive an exact optimal test plan using integer programming, but this solution approach is {\em NP} hard \cite{schrijver1998theory}. We have developed a less computationally expensive approach by treating the number of component tests as non-negative real numbers rather than discrete integers. 

In our alternative solution method, we maximize $N_{min}$ in the continuous domain using linear programming, then convert the continuous test plan values back to discrete integers. The approach is described in more detail in the section below, and an example R script implementation of the method is given in Appendix \ref{sec:appendix}.

\section{Test Plan Optimization using Linear Programming}
Let us denote 
\begin{itemize}
	\item[$m$] is the number of components;
	\item[$\mathbf{f}$] $= (f_1, f_2, \ldots, f_m)' \in \mathbb{R}^m$ is the fraction of tests allocated to each component, i.e. $f_j = n_j/N,~j=1..m$;
	\item[$s$]  is the number of minimal cutsets
	\item[$\mathbf{1}_s$] $ = (1, 1, \ldots, 1)'$ is a unit vector of size $s$	
	\item[$Y$]  is a $ s \times m $ incidence matrix for minimal cutsets where
	$y_{ij} = 1$ if component $c_j$ belongs to minimal cutset $i$, $y_{ij} = 0$ otherwise.   
\end{itemize}

In order to maximize the minimum number of tests across all minimal cutsets, we are looking for the best among (sub-optimal) test plans that allocate the same fraction of tests $g$ to all minimal cutsets in $Y$, by solving the following linear programming (LP) problem:
\begin{eqnarray}
&& \label{LPP1} g \rightarrow \max \\
& \textrm{given}& \nonumber \\
&& Y \cdot \mathbf{f} = g.\mathbf{1}_s; \label{eqn:sumfrac}\\
&& \sum_{j=1}^m f_j = 1;\\
&& f_j \ge 0,~j=1..m,
\end{eqnarray}
where $Y \cdot \mathbf{f}$ is the matrix product of a matrix and a vector that computes sum of the component test fractions for every cutset, hence constraint (\ref{eqn:sumfrac}) requires that $\sum_j (y_{ij}.f_j)=g,~~ i=1..s$. 

We can now eliminate variable $g$ by defining the following terms:
\begin{eqnarray}
&& \mathbf{h} = \mathbf{f}/g\\
&& H = 1/g.
\end{eqnarray}

Rewriting the LP problem in these terms, $g$ is maximized when $H$ is minimized, i.e.:
\begin{eqnarray}
&& \sum_{j=1}^m h_j = H \rightarrow \min \\
& \textrm{given} & \nonumber\\
&& \label{constr1} Y \cdot \mathbf{h} = \mathbf{1}_s; \\
&& h_j \ge 0,~j=1..m.
\end{eqnarray}

The R $simplex()$ LP solver function can be used to derive the solution to this problem. In practice however, this function can sometimes fail to find a solution when  equality constraints are used -- probably because it fails to generate an initial feasible point. To resolve the issue, we noted that $H$ reaches its unconstrained minimum when $h_j = 0,~j=1..m$. Therefore, equality constraint (\ref{constr1}) can be replaced with the inequality constraint $Y \cdot \mathbf{h} \ge \mathbf{1}$, resulting in the following LP problem
\begin{eqnarray}
&& \sum_j h_j = H \rightarrow \min \\
& \textrm{given} & \nonumber \\
&& Y \cdot \mathbf{h} \ge \mathbf{1}_s; \\
&& h_j \ge 0,~j=1..m.
\end{eqnarray}
This optimization problem can solved with an R script that calls the LP solver $simplex()$ as shown in Appendix~\ref{sec:appendix}.

The resultant optimal test allocation fractions for the components are: 
\begin{eqnarray}
&&\mathbf{f}_{op}=\mathbf{h}_{op}/H_{op}
\end{eqnarray}
and the optimal minimal cutset fraction $g_{op}$ is: 
\begin{eqnarray}
&&g_{op}=1/H_{op}.
\end{eqnarray}
As in general these fractions are continuous real values, the optimal apportionment of component tests i.e. $\mathbf{n}=\mathbf{f}_{op}N$ can be non-integer. An optimal integer component test allocation can be derived by 
first finding the smallest test multiple, $N_{0}$, where all component test fractions scale to integer values, i.e.
\begin{eqnarray}
\lfloor \mathbf{f}_{op}N_{0}\rfloor=\mathbf{f}_{op}N_{0}.
\end{eqnarray}

$N_{0}$ can be found by incrementing an integer number $k$ by 1 until all the products $k \cdot f_j , ~ j = 1..m$ become integer.

The optimal plan for a total number of tests
\begin{equation}
N^- = N - (N~mod~N_{0})
\end{equation}
is always integer. The remaining $(N~mod~N_{0})$ tests 
can be allocated arbitrarily to any of the components (or not allocated at all) because they cannot increase the value of $N_{min}$.

If there is an option to add small number of tests to the plan, one can consider a test plan for $N^+$ tests where
\begin{equation}
N^+ = N^- + N_{0}.
\end{equation}
 
\section {Example}

\begin{figure}[h]
	\begin{center}
		\includegraphics[width=0.7\textwidth]{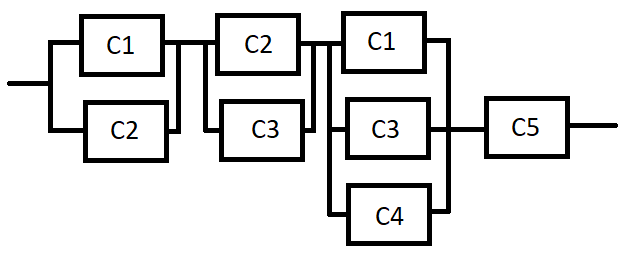}
	\end{center}
	\caption{Example asymmetric RBD}
	\label{fig: RBD1} 
\end{figure}

Let us consider an example asymmetric structure with the reliability block diagram (RBD) given in Figure \ref{fig: RBD1}.

Its minimal cutsets are: 
\begin{eqnarray}
&& C1, C2 \nonumber\\
&& C2, C3 \nonumber\\
&& C1, C3, C4 \nonumber\\
&& C5 
\end{eqnarray}
and its minimal cutset matrix $Y$ is shown in Table \ref{tbl:incidence_matrix}.

\begin{table}[ht]
\caption{Minmal cutset incidence matrix}
\label{tbl:incidence_matrix}
\begin{center}
\begin{tabular}{| c | c c c c c |}
	\hline
	 cutset & \multicolumn{5}{|c|}{component j} \\
	i  &  1 &  2 &  3 &  4 &  5 \\
	  \hline
	1 &  1 &  1 &  0 &  0 &  0 \\
	2 &  0 &  1 &  1 &  0 &  0 \\
	3 &  1 &  0 &  1 &  1 &  0 \\
	4 &  0 &  0 &  0 &  0 &  1 \\
	\hline
\end{tabular}

\end{center}
\end{table}

For this minimal cutset incidence matrix,
the R script generates the following optimal test allocation fractions:

\begin{center}
\begin{tabular}{c c c c c c}
\hline	
$f_1$  &  $f_2$ &  $f_3$  &   $f_4$ &  $f_5$ & $g_{op}$ \\ 
\hline
0.2 & 0.2 & 0.2 & 0.0 & 0.4 & 0.4\\ 
\hline 
\end{tabular}
\end{center}
where zero tests are allocated to component $c_4$.

For this plan, sequential search gives $N_{0} = 5$. Therefore,
for a test campaign with a total number of tests, $N=20003$, we have
\begin{equation}
N^- = 20003 - (20003~mod~5) = 20000
\end{equation}
with the test allocation 
\begin{center}
	\begin{tabular}{c c c c c c}
		\hline	
		$n_1$  &  $n_2$ &  $n_3$  &   $n_4$ &  $n_5$ & $N^-$ \\ 
		\hline
		4000 & 4000 & 4000 &   0 & 8000 & 20000\\ 
		
		\hline 
	\end{tabular}
\end{center}
and the least number of tests allocated to any minimal cutset is $N_{min}= g_{op}\cdot N^- = 8000$. 

By comparison, if we use the strategy proposed in \cite{bishop2020conservative} of allocating $N/P$ tests equally to components on a single shortest success path, such as $(c_1, c_2, c_5)$, then $P=3$. This is clearly sub-optimal as the least tested cutsets only have $\lfloor N/P \rfloor = N_{min}=6667$ tests. 

\section{Concluding Remarks}

It can be observed that the fractions generated in the continuous domain are independent of the number of \-tests, so they only need to be generated once for any given structure. It is only the integer test plan that needs to be recalculated for a given test budget -- reducing the computing resources needed for a new plan.

In principle, it would be possible to create a library of optimal test plan solutions for different structures that can be converted to integer test plans for any specified number of component tests.

This strategy of solving in the continuous domain and then efficiently deriving optimal (or near optimal) solutions in the integer domain might be applicable to other problem areas.

\appendix 

\section{Test Plan Optimization R Script}\label {sec:appendix}

The test plan optimization approach was implemented using the standard simplex solver available in the R statistical analysis library. 

The use of the test plan optimizer is illustrated using non-symmetric structure shown in Figure~\ref{fig: RBD1}.

{\normalsize 
\begin{Verbatim}[tabsize=4]
library("boot")

#------------------------------------------
# lptplan_example <- function( N, alpha)
# N - total number of tests (default 20003)
# alpha = 1 - confidence level (default 0.05)
#------------------------------------------ 

lptplan_example <- function(
N=20003, 
alpha = 0.05
)
{
# minimal cutset matrix
	cutsets <- matrix(
	c(
	1,1,0,0,0,  # cutset: C1, C2
	0,1,1,0,0,  # cutset: C2, C3 
	1,0,1,1,0,  # cutset: C1, C3, C4 
	0,0,0,0,1   # cutset: C5
	), 4, 5, 
	byrow=TRUE
	)
	
# Generate optimized test plan
	print  ( lptestplan(cutsets, N, alpha) )
}

#-----------------------------------------
# lptestplan <- function(cutsets, N, alpha)
# cutsets 
#  incidence matrix for the minimal cutsets
#  columns represent components
#  rows represent cutsets
# N   total number of tests
# alpha = 1 - confidence level
#-----------------------------------------     

lptestplan <- function(cutsets, N, alpha)
{   
# Number of components
	m <- ncol(cutsets)
	
# Number of minimal cutsets
	s <- nrow(cutsets)

# Unit vectors
	uvm <- rep(1,m)
	uvs <- rep(1,s)
	
# Solve LP
	lp0 <- simplex(
		a = uvm, 
		A3 = cutsets, 
		b3 = uvs
	)
	H = as.numeric(lp0$value)
	h = lp0$soln
	
# Optimal cutset test fraction 
	g <- 1/H
	
# Optimal component test fractions
	f <- h * g
		
# Find minimal integer test plan
	k <- 1
	r <- 1
	while(r>0){
		r <- sum ((f*k)%%1)
		if(r>0) k <- k+1
	}
	N0 <- k
	N_minus <- N - (N%%N0)
	
# Generate integer test plan
	N_min <- N_minus * g
	lptest_plan <- N_minus * f
	
# Calculate upper confidence bound
	q_u <- log(1/alpha)/N_min
	
# Return optimized result
  return
  (
	list(
	cutsets=cutsets,
	alpha = alpha,
	component_fractions = f,
	cutset_fraction = g,
	N = N,
	N0 = N0,
	N_minus = N_minus,
	lptest_plan = lptest_plan,
	N_min = N_min,
	q_u = q_u 
	)
  )
}
\end{Verbatim}
}

\bibliography{component_test_optimisation_v1}
\bibliographystyle{ieeetr}

\end{document}